\documentclass[conference]{IEEEtran}
\usepackage{cite}
\usepackage{amsmath,amssymb,amsfonts}
\usepackage{algorithmic}
\usepackage{graphicx}
\usepackage{textcomp}
\usepackage{xcolor}
\usepackage{url}
\usepackage{adjustbox} 
\usepackage{longtable} 
\usepackage{booktabs}
\usepackage{array} 
\usepackage{caption}
\usepackage{tikz}
\usepackage{xcolor}
\usepackage{fontawesome5}
\usepackage{multicol}
\usepackage{hyperref}
\def\BibTeX{{\rm B\kern-.05em{\sc i\kern-.025em b}\kern-.08em
 T\kern-.1667em\lower.7ex\hbox{E}\kern-.125emX}}

\begin{document}

\title{Digital Transformation in the Petrochemical Industry: Challenges and Opportunities in the Implementation of IoT Technologies}

\author{
    \IEEEauthorblockN{Noel Portillo \href{https://orcid.org/0009-0009-8817-3187}{(ORCID: 0009-0009-8817-3187)}}
    \IEEEauthorblockA{        
        \textit{Axxis Technologies} \\
        Houston, TX, USA \\
        noel@axxistechnologies.com
    }
    \and
    \IEEEauthorblockA{
        \textit{Independent Researcher} \\
        Ahuachapán, El Salvador \\
        noel.portillo@ieee.com
    }
}

\maketitle

\begin{abstract}
The petrochemical industry faces significant technological, environmental, occupational safety, and financial challenges. Since its emergence in the 1920s, technologies that were once innovative have now become obsolete. However, factors such as the protection of trade secrets in industrial processes, limited budgets for research and development, doubts about the reliability of new technologies, and resistance to change from decision-makers have hindered the adoption of new approaches, such as the use of IoT devices.
This paper addresses the challenges and opportunities presented by the research, development, and implementation of these technologies in the industry. It also analyzes the investment in research and development made by companies in the sector in recent years and provides a review of current research and implementations related to Industry 4.0.
\end{abstract}


\section{Introduction}
In recent years, a large number of IoT devices and software have been developed, capable of processing vast amounts of information and supporting critical processes. For example, Al-Nahrain University, in its Iraqi Journal of Information and Communication Technology, published an article emphasizing the early detection of breaches in pipeline monitoring systems, a crucial concern for oil management companies. The article highlights how various technologies have been developed in recent years to identify pipeline leaks and presents a solution based on IoT devices such as the Arduino Mega and the BLYNK IoT platform \cite{b1}. These next-generation devices can replace obsolete control systems still used in most industries, particularly in petrochemicals. Elete \cite{b2}, in their 2024 article, provides a comprehensive approach by studying successful case studies of testing in production environments across upstream, midstream, and downstream operations, highlighting the significant role of digital transformation in driving innovation.

This paper begins with a review of the history and evolution of technology adoption in the industry, as well as the current and future challenges it faces. It analyzes the opportunities presented by replacing mechanical and analog devices with IoT devices for monitoring, control, and supervision of industrial processes, as shown by Belyaev\cite {b3} in their research on developing a methodological framework for building existing automated process control systems (ATPCS). Additionally, the paper investigates market participation, profitability, and research and development investment in the sector's companies, as discussed by Patterson\cite{b4}. It also references the market share of major players selling automation technology solutions to petrochemical plants in the United States. Finally, the paper addresses and describes in detail the challenges faced by this industry, exploring topics such as industrial safety, government regulations, and technological standards that must be considered to develop IoT-based automation systems for the industry.

The main objective of this article is to explore the historical origins of automation in the petrochemical industry, analyze its technological evolution over time, and examine the current challenges in adopting advanced technologies associated with Industry 4.0.

\section{Brief History of Automation}
The exact moment when the first machines emerged is unknown, but humanity has been using methods to facilitate work and improve the quality of life since ancient times. The earliest records of machine use date back to Mesopotamia between 4000 and 2900 BCE. Artifacts discovered from that period include what are now known as the six simple machines: the wedge, the inclined plane, the lever, the wheel and axle, the screw, and the pulley, as described by Faiella \cite{b5} and Moorey \cite{b6}.
In modern times, in 1785, American engineer Oliver Evans invented the flour mill, which can be considered the first fully mechanical machine. The system consisted of bucket elevators, conveyor belts, and Archimedean screws that interacted with the motion of a wheel powered by hydraulic energy. This invention revolutionized the grain processing industry of the time and laid the foundation for the development of new automatic machines\cite{b7}.
\begin{figure}[ht!]
    \centering
    \includegraphics[width=0.45\textwidth]{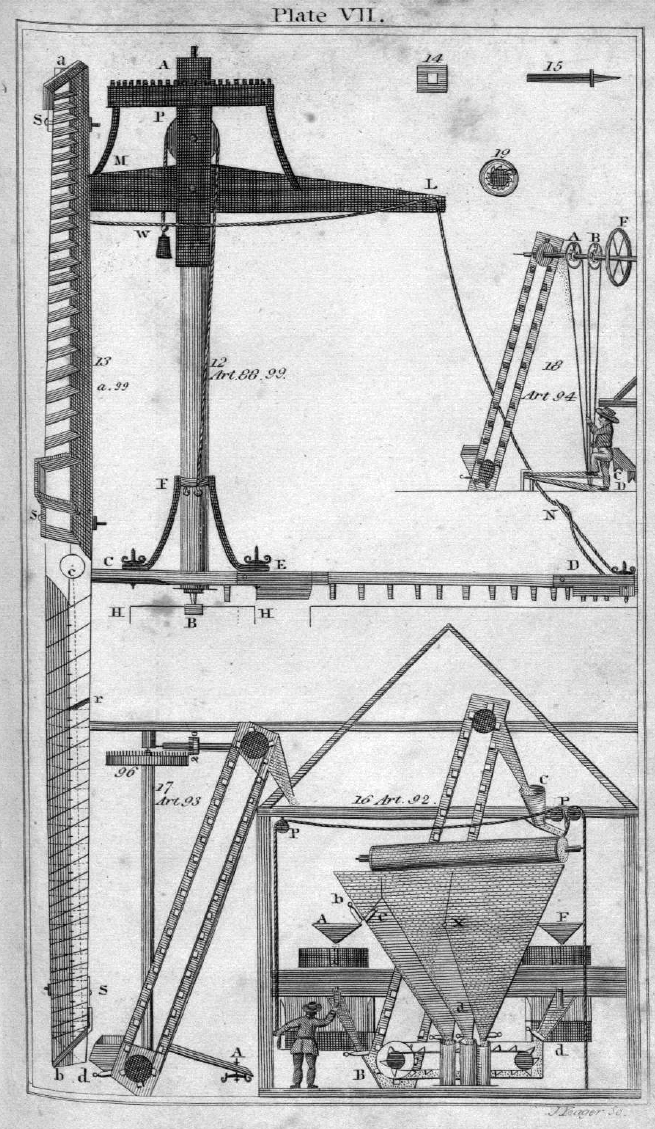}
    \caption{Flour Mill (Oliver Evans) and the Hopper Boy System. Source: Wikimedia Commons.}
    \label{Hopper Boy System}
{\footnotesize Source: \cite{b10}.}
\end{figure}

In 1835, the American physicist Joseph Henry invented the first electromechanical relay, a device that allows the activation of an electrical circuit from a distance using another circuit with a weaker current. This invention laid the foundation for what is known as relay logic, which began to be used in industry in 1920 for automatic control (Wikipedia\cite{b9})

\begin{figure}[ht!]
    \centering
    \includegraphics[width=0.45\textwidth]{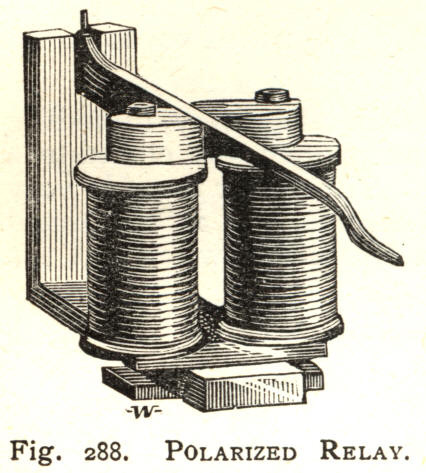}
    \caption{First Magnetic Relay (Joseph Henry). }
    \label{First Magnetic Relay}
{\footnotesize Source: \cite{b10}.}
\end{figure}

In 1947, a milestone in modern electronics was achieved with the invention of the transistor by American engineers John Bardeen, Walter Brattain, and William Shockley at Bell Labs. The transistor replaced vacuum tubes, which were analog devices that were inefficient in many aspects. This invention laid the foundation for modern electronics, and together with other advances such as the incorporation of new materials, techniques, and design and manufacturing processes, it led to the development of the first microprocessor in 1971. This microprocessor was based precisely on the reduction in size of transistors and integrated circuits.

Although Programmable Logic Controllers (PLCs) had already been developed in 1969 by Modicon, they did not yet include a microprocessor. In 1973, the same company created the first PLC using an 8-bit microprocessor, the Modicon 184. These devices were quickly adopted by many industries to make their production machines more efficient.

\section{The Industrial Revolution and the Foundations of the Petrochemical Industry}

The phenomenon known as the Industrial Revolution began in the mid-18th century, originating in Great Britain and later spreading to other parts of the world\cite{b11}. During this period, new methods and technologies emerged that accelerated production processes, including the use of the steam engine. This innovation provided society not only with a means of transportation but also the ability to construct machinery that greatly facilitated industrial production. These revolutionary manufacturing methods increased the demand for energy sources, particularly coal and hydrocarbons. Additionally, advances in chemistry enabled the development of new production methods, leading to the creation of materials such as the first polymers and other petroleum-derived byproducts.

As previously described, the use of automatic controls in industry, specifically in petrochemicals, had already been adopted in many areas. However, at that time, partially automated controls based on relay logic were used. It was not until the 1980s that the use of Programmable Logic Controllers (PLCs) in refinery processes became widespread. The need to automate potentially hazardous processes, characteristic of this industry, led to the adoption of these systems for monitoring and controlling critical and high-risk processes. This also resulted in increased productivity and the expansion of oil exploitation worldwide.

\section{Current State of Automation in the Petrochemical Sector}
The process of oil refining involves a vast number of people, processes (both automated and manual), materials, contracting companies, and an extensive amount of administrative management. This ranges from public relations and dealings with regulatory and governmental agencies to handling lawsuits and protecting trade secrets, resulting in unprecedented complexity within this sector. It is no surprise that changes and the adoption of new technologies and processes are difficult to implement. Additionally, there are no specific studies on the level of technological adoption within this industry. What we know comes primarily from what organizations publish through their public relations and marketing offices. Therefore, the best way to understand the current state of automation in the industry is through the contractors who work for these companies and have direct access to the refineries.

According to the website offshore-technology.com\cite{b12}, citing data from GlobalData's Petrochemicals Database, there are 385 active refinery complexes in North America, with 67 more planned. Of the top 10 complexes, six are located in Texas, and the remaining four are in Louisiana, a neighboring state. The newest of these complexes has been in operation for 38 years. The byproducts produced in these refineries are primarily used in the manufacturing of plastics, insulation materials like polyurethane, synthetic fibers, polyester, cosmetics, and pharmaceutical products. These chemicals are essential for the consumer goods, infrastructure, and technology industries.

\begin{table}[h!]
    \centering    
    \begin{adjustbox}{max width=\columnwidth}
    \begin{tabular}{@{}p{0.6cm}p{2.5cm}p{1.8cm}p{1.5cm}p{2.8cm}@{}}
        \toprule
        \textbf{Pos.} & \textbf{Name} & \textbf{Location} & \textbf{Year} & \textbf{Products} \\ \midrule
        1  & Formosa Plastics Group Point Comfort Complex & Texas & 1978 & Ethylene, polyethylene, EDC, EG, polypropylene \\
        2  & LyondellBasell Channelview Complex & Texas & 1957 & Ethylene, propylene, ethylbenzene, styrene, propylene oxide \\
        3  & Exxon Mobil Corporation Baytown Complex & Texas & 1940 & Ethylene, propylene, polypropylene, xylene \\
        4  & Dow Freeport Complex & Texas & 1951 & Ethylene, propylene, polyethylene, PO \\
        5  & CF Industries Donaldsonville Complex & Louisiana & 1966 & Ammonia, urea \\
        6  & Exxon Mobil Corporation Baton Rouge Complex & Louisiana & 1942 & Polyethylene, ethylene, propylene, polypropylene \\
        7  & Occidental Chemical Corporation Ingleside Complex & Texas & 1989 & EDC, vinyl chloride monomer (VCM) \\
        8  & Westlake Chemical Corporation Lake Charles Complex & Louisiana & 1986 & EDC, ethylene, VCM, polyethylene \\
        9  & Dow Plaquemine Complex & Louisiana  & 1958 & Polyethylene, ethylene, propylene, PO, EG \\
        10 & Chevron Phillips Chemical Company Baytown Complex & Texas & 1963 & Ethylene, polyethylene, propylene \\
        \bottomrule
    \end{tabular}
    \end{adjustbox}
 \caption{Top 10 Petrochemical Complexes in North America}
    \label{tab:petrochemical_complexes}
{\footnotesize Source: \cite{b13}.}
\end{table}

\section{Opportunity for Change}
In the United States, there is a growing sector of companies offering a wide range of services to the petrochemical industry, including engineering, excavation, mechanical engineering, machinery maintenance, cleaning, and catalytic product treatment, among others. These companies are adopting new technologies to carry out projects in a more cost-effective, safer, and more efficient manner. They represent a significant opportunity for researchers, as they have direct access to industrial complexes. For this reason, these companies are emerging as key players in the development of IoT-based solutions. Their deep understanding of processes, combined with the need for innovation, creates an ideal environment for research, testing, and the development of technological solutions that are safe, cost-effective, and sustainable.

\section{Industry 4.0}
The concept of Industry 4.0 was first introduced at the Hannover Fair in Germany in 2011. The Fourth Industrial Revolution, or Industry 4.0 as it is commonly known today, introduces a series of advanced technological solutions designed to create smart factories. This paradigm marks a significant shift from the use of traditional technologies to disruptive technologies. These include Artificial Intelligence (AI), the Internet of Things (IoT), Cloud Computing, Big Data, Machine Learning, and Data Science, among other emerging technologies that integrate both vertical and horizontal solutions throughout the entire production chain.

A study conducted by the firm Mordor Intelligence\cite{b14} indicates that the North American market leads in the adoption of new technologies, particularly in sectors like the petrochemical industry. According to the report, this sector is expected to achieve a compound annual growth rate (CAGR) of 9.50\%\ between 2019 and 2029, using 2023 as the base year.

An interesting aspect highlighted in the report is the participation of major tech giants as key players in the implementation of these technologies, as illustrated in the following graph:

\begin{figure}[h!]
    \centering
    \includegraphics[width=0.5\textwidth]{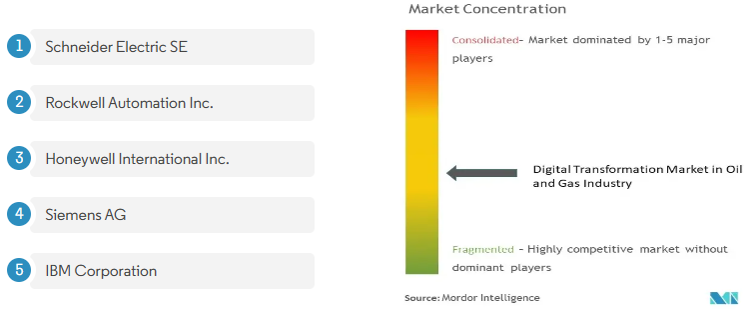}
    \caption{Market Concentration of Disruptive Technologies. Source: Mordor Intelligence.}
    \label{concentration_market} 
{\footnotesize Source: \cite{b14}.}
\end{figure}

The graph highlights several key points about the current state of disruptive technology adoption in the petrochemical industry. First, it is evident that large corporations continue to lead in this field. This is not surprising, given that they possess the expertise, infrastructure, and research and development (R\&D) budgets necessary to compete in highly specialized sectors like this one. These advantages enable these companies to secure significant contracts with industries such as petrochemicals and manufacturing.  
However, these same characteristics can pose a challenge to the widespread adoption of new technologies in certain cases. Two main factors contribute to this phenomenon:

\begin{enumerate}
    \item \textbf{High implementation and maintenance costs:} The technological solutions offered by large corporations are often extremely expensive, limiting their accessibility to small and medium-sized enterprises.
    \item \textbf{Proprietary technologies:} Many of these solutions are closed, meaning there is no public documentation or open standards to enable easy integration or independent development.
\end{enumerate}

\subsection{Opportunities for startups and individuals}
Although large corporations currently dominate this market, these limitations also create an opportunity for startups and innovative individuals to enter the industry. By developing more accessible solutions based on open standards and open-source technologies, these entities can meet the needs of small and medium-sized enterprises seeking to modernize their operations. 

However, this topic warrants a deeper analysis from another perspective, which we will not address here.  

Industry 4.0 continues to transform key sectors like petrochemicals but faces challenges related to accessibility and reliance on large corporations. This raises questions about how to democratize the adoption of disruptive technologies and open the market to new, more flexible, and accessible solutions.

\section{Challenges for the implementation of IoT solutions in the industry.}

As mentioned earlier, the implementation of IoT solutions in the petrochemical industry faces various challenges, including industrial safety, economic resources for R\&D, and trade secrets that even reach the level of national security and are highly regulated by governments. Let us explore these three aspects to better understand them.

\subsection{Industrial Safety}

In the United States, the main agency regulating industrial safety in chemical processing plants for petroleum and natural gas derivatives is OSHA (Occupational Safety and Health Administration), an agency of the U.S. Department of Labor (DOL). 

This agency has developed strict regulations and a framework that must be followed in the operation of petrochemical plants. This is not an exaggeration as previously mentioned, the implementation of IoT solutions in the petrochemical industry faces significant challenges, including compliance with industrial safety regulations. The following notable incidents emphasize the importance of adhering to these regulations:

\begin{itemize}
    \item In 1984, Bhopal, India, a leak of the chemical methyl isocyanate occurred at a pesticide factory owned by the American company Union Carbide Corporation. This incident resulted in the death of between 15,000 and 20,000 people, with half a million survivors exposed to the contamination\cite{b15}.
    \item In October 1989, at the Phillips Petroleum Company plant in Pasadena, TX, 23 workers lost their lives, and 130 others were injured during an explosion within the facility. It is estimated that metal debris from the explosion reached up to six miles around the complex\cite{b16}.
    \item On October 10, 2024, a gas leak at the Pemex refinery in Deer Park, TX, claimed the lives of three workers and left 13 more hospitalized\cite{b17}.
\end{itemize}

The most significant regulatory framework for industrial safety in refineries is the \textit{Process Safety Management (PSM)} standard\cite{b18}, a document that meticulously outlines all safety measures required for operations. Below, we focus on the requirements for the use of electrical and electronic equipment under this standard.

\subsubsection{Information on Equipment Used in the Process}

The PSM standard requires that information about the equipment in the process include the following:

\begin{itemize}
    \item Materials of construction,
    \item Piping and instrumentation diagrams (P\&ID),
    \item Electrical classifications,
    \item Relief system design and design basis,
    \item Ventilation system design,
    \item Design codes and standards employed,
    \item Material and energy balances for processes constructed after May 26, 1992, and
    \item Safety systems (e.g., interlocks, detection, or suppression systems).
\end{itemize}

Employers must document that the equipment complies with generally recognized and accepted engineering practices (GRAEP). For existing equipment designed and built according to codes, standards, or practices no longer in widespread use, employers must determine and document that the equipment is designed, maintained, inspected, tested, and operated safely.

\subsubsection{IoT Devices and Industrial Safety}

As we can see from the PSM standard, electrical and electronic devices must be properly documented. However, it is important to highlight that, from the perspective of industrial safety, the standard does not explicitly prohibit or exclude the use of new technologies such as IoT. Furthermore, there are various modern standards ensuring the physical robustness of IoT devices for highly demanding industrial environments, including:

\begin{itemize}
    \item ATEX\cite{b19},
    \item IECEx\cite{b20},
    \item UL 1203\cite{b21},
    \item Protection classifications such as IP\cite{b22}, and
    \item NEMA\cite{b23}.
\end{itemize}

\subsection{Industrial Trade Secrets}

The energy and industrial sectors represent strategic resources worldwide. For this reason, strict measures are adopted to protect trade secrets related to production techniques and equipment design. This not only prevents the possibility of such secrets being copied but also mitigates the risks associated with leaks of sensitive information that could compromise national security. In this context, efforts have been made to develop advanced techniques, such as offline authentication methods for industrial control software based on national secrets\cite{b24}. 

It is not surprising that these industries are considered of national interest and are protected by rigorous regulations established by governments. Consequently, physical or digital access to critical resources, such as operating manuals, operational metrics, formulas, and designs, is often regulated or restricted. This is especially true when dealing with confidential or highly sensitive information for industry and national security.

For researchers and students, access to industrial environments where petroleum and gas are refined and transformed into chemical by-products is limited due to strict industrial safety protocols and confidentiality policies. These measures aim to protect both the integrity of the processes and the safety of people and the environment. However, some companies and educational institutions promote collaboration through internships, supervised visits, and joint research projects. These initiatives enable a controlled transfer of knowledge.

In this context, it is important to highlight that the lack of practical experience in these environments hinders the ability to propose applicable solutions for industrial production. Companies leading the implementation of technologies in this sector are often those with prior experience in these areas, giving them a significant competitive advantage. In contrast, industries such as mass consumption, household products, home automation, and communications evolve and adopt disruptive technologies much faster. This is because researchers, inventors, and developers have greater access to testing environments and can prototype and implement solutions more agilely.

Another critical concern in the implementation of IoT technology is that these devices often rely on internet connectivity for communication, remote programming, and real-time monitoring. This exposes them to significant risks of cyberattacks. In 2017, the first cyberattack specifically designed to harm human targets occurred in a petrochemical plant in Saudi Arabia\cite{b25}. The ransomware known as Triton, named after the Schneider Electric Triconex safety system it targeted, managed to breach the industrial complex's last line of defense, taking control of its safety systems\cite{b26}. Although the attack failed due to a flaw in its code, it raised global alerts about the security of internet-connected devices in industrial environments.

Currently, a vast number of sensors, valves, actuators, and PLCs are internet-enabled, allowing for remote control and monitoring. This reality has increased concerns regarding the necessary security measures, representing a considerable challenge for the implementation of technologies associated with Industry 4.0. For this reason, companies in critical sectors must carefully evaluate these risks before undertaking technological transitions.

The industry is now taking action to address the risks associated with using disruptive technologies in industrial complexes. Increasingly, regulations are being developed to establish the necessary safety measures before implementing Industry 4.0 projects in critical sectors such as the petrochemical industry.

\begin{table}[ht]
\centering
\setlength{\tabcolsep}{4pt} % Reduce el espacio entre columnas
\renewcommand{\arraystretch}{1.3} % Ajusta la altura de las filas
\begin{tabular}{|m{0.25\linewidth}|m{0.25\linewidth}|m{0.4\linewidth}|}
\hline
\textbf{Standard} & \textbf{Institution} & \textbf{Application} \\ \hline
ISA/IEC 62443 & International Society of Automation (ISA) & 
\begin{itemize}
    \item This set of standards focuses on the security of industrial automation and control systems (ICS), including IoT devices.
    \item Defines requirements for manufacturers, integrators, and operators, covering topics such as secure design, risk assessment, and protection against cyberattacks.
\end{itemize} \\ \hline
NIST Cybersecurity Framework (CSF) & National Institute of Standards and Technology (NIST) & 
\begin{itemize}
    \item Published by the U.S. National Institute of Standards and Technology (NIST), it provides a framework for managing cybersecurity risks, including IoT devices.
    \item Includes specific guidelines to protect industrial systems operating with connected devices.
\end{itemize} \\ \hline
NIST SP 800-183 (IoT Trustworthiness) & National Institute of Standards and Technology (NIST) & 
Provides guidelines to ensure IoT devices are reliable and secure in industrial environments, emphasizing aspects such as privacy, authentication, and access control. \\ \hline
API RP 1164 & American Petroleum Institute & 
Developed by the American Petroleum Institute (API) for the cybersecurity of control systems in pipelines and petrochemical plants, including industrial IoT. \\ \hline
ISO/IEC 27001 and 27002 & International Organization for Standardization (ISO) and the International Electrotechnical Commission (IEC) & 
International standards that establish requirements for information security management systems, essential for protecting networks and IoT devices in industrial environments. \\ \hline
IEEE P2413 & Institute of Electrical and Electronics Engineers & 
An architectural framework for IoT addressing interoperability, security, and privacy in connected devices, including industrial applications. \\ \hline
\end{tabular}
\caption{Applicable regulations for cybersecurity in industrial complexes}
\label{table:cybersecurity-standards}
\end{table}

\subsection{Economic Resources (R\&D)}

The petrochemical industry is one of the most powerful sectors in the world in terms of revenue. Between 2021 and 2023, the four major refinery giants in the United States—Exxon Mobil, Shell, Total Energies, and Chevron—reported estimated profits of \$332.6 billion. This was largely driven by an increase in oil prices, influenced by production cuts from OPEC and other factors affecting crude oil prices\cite{b27}.

\begin{figure}[ht!]
    \centering
    \includegraphics[width=0.5\textwidth]{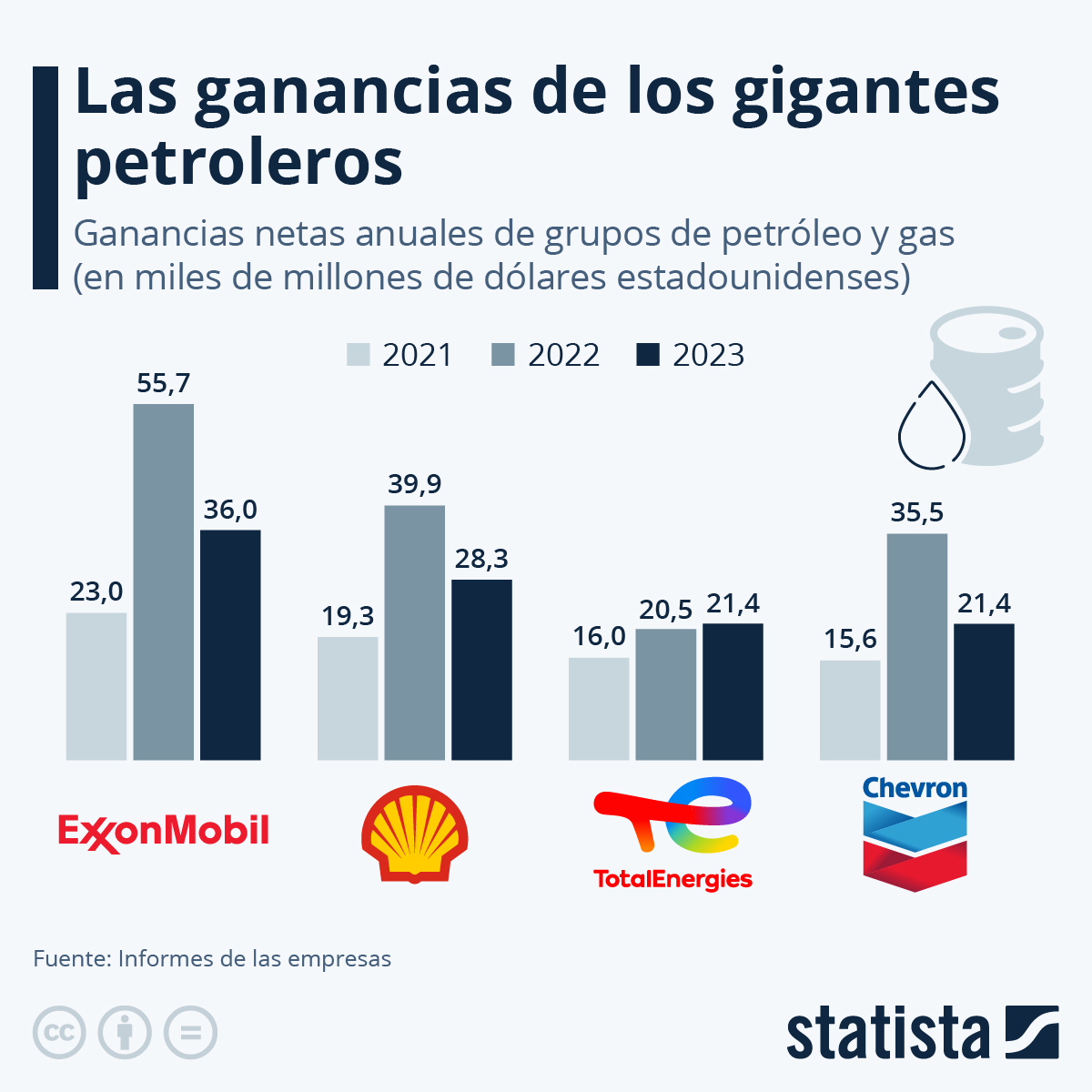}
    \caption{Annual Net Profits.}
    \label{anual_profits}
{\footnotesize Source: \cite{b28}.} 
\end{figure}

However, regarding R\&D in the sector, specifically in the development and implementation of new technologies for control, monitoring, prevention, and data collection, the investment by these companies is minimal compared to their annual revenues:

\begin{figure}[h!]
    \centering
    \includegraphics[width=0.5\textwidth]{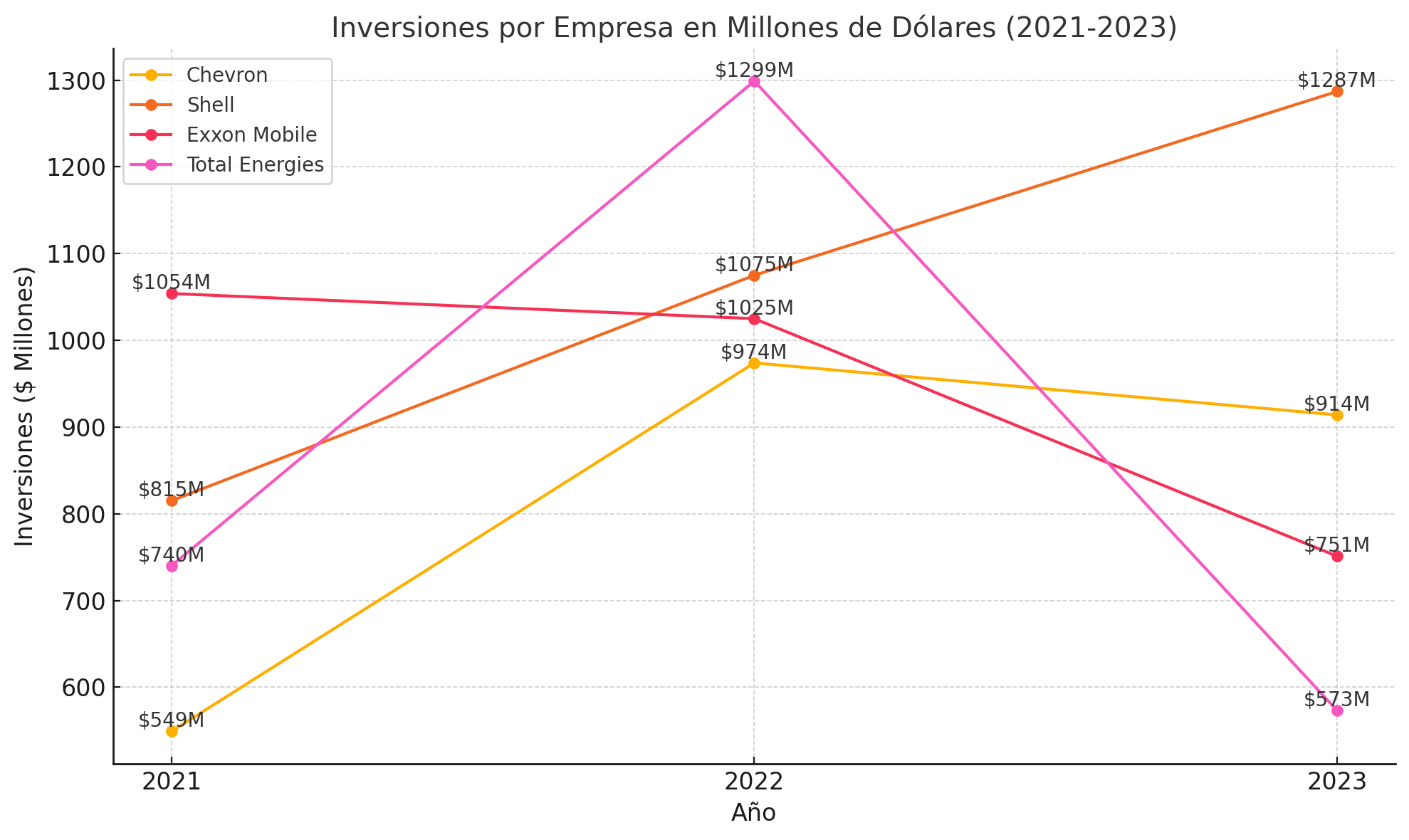}
    \caption{Investment in Research and Development.}
    \label{investment-vs-profits} 
{\footnotesize Source: \cite{b28}.}
\end{figure}

\section{Conclusions}

The implementation of IoT technologies in the petrochemical industry holds enormous potential to optimize critical processes, reduce operational costs, and improve industrial safety. However, this sector faces significant challenges, such as resistance to change, high initial costs, and cybersecurity concerns.

A key aspect to overcoming these barriers is collaboration between researchers, developers, and the academic sector. Working alongside contracting companies with direct access to industrial complexes facilitates the study and testing of disruptive technologies. This not only benefits technology companies but also contractors by generating more efficient solutions for their operations.

The integration of IoT and other disruptive technologies is transforming traditional industrial automation systems as well as other sectors, such as consumer goods. In the petrochemical industry, large corporations are leading this change, but the democratization of these technologies could open opportunities for small and medium-sized enterprises. For these new players, it is essential to comply with industrial and cybersecurity regulations, such as the ISA/IEC 62443 and ATEX standards, which provide robust frameworks to ensure reliability and safety in complex environments.

On the other hand, although the petrochemical industry generates significant revenue, investment in research and development remains limited compared to other sectors. This slows the adoption of innovative technologies. However, sharing success stories and raising awareness of the benefits of technological change can influence industry leaders to take a more proactive approach to these new opportunities.

\section{Recommendations}

Petrochemical companies must prioritize investment in research and development (R\&D), not only to remain competitive in a constantly evolving global market but also because these technologies have proven effective in monitoring critical processes that currently pose environmental challenges. A clear example is the monitoring of greenhouse gas leaks in wells and refineries, as discussed by Arinze\cite{b29}.

In addition, it is essential to implement robust cybersecurity strategies, including the adoption of international standards and continuous monitoring of IoT systems. This is particularly important considering that IoT devices use a wide range of communication protocols, such as Wi-Fi, Bluetooth, and radio frequency, which are vulnerable to cyberattacks.

Another crucial aspect is promoting the use of open technologies and standards, as these can reduce costs and facilitate the integration of new solutions into industrial environments. This would enable more companies, including small and medium-sized enterprises, to access these innovations. At the same time, it is vital to encourage training and cultural change within organizations by establishing education and awareness programs for decision-makers and technical staff. These initiatives will help reduce resistance to change and promote a culture focused on technological innovation.

Finally, establishing strategic alliances between governments, educational institutions, and companies is a key factor in accelerating the adoption of IoT technologies. Such alliances will ensure that these technologies are implemented safely and effectively.

All these actions will enable the petrochemical industry to evolve, become more profitable, improve the quality of life of citizens, and, most importantly, significantly reduce financial, labor, and environmental risks.


\begin{thebibliography}{00}
\bibitem{b1} "Oil and Gas pipelines monitoring using IoT platform," *Iraqi Journal of Information and Communication Technology*, vol. 6, no. 1, pp. 9-27, 2023. doi: 10.31987/ijict.6.1.209.
\bibitem{b2} Y. E. Tari, E. O. Nwulu, V. E. Erhueh, O. A. Akano, and T. A. Aderamo, "Exploring advanced techniques in process automation and control: A generic framework for oil and gas industry applications," *ESTJ*, vol. 5, no. 11, 2024. doi: 10.51594/estj.v5i11.1704.
\bibitem{b3}P. S. Belyaev, W.-T. Khu, L. G. Varepo, G. Berdaliyeva, A. Ussenova, and B. Artykbay, "Flexible control systems in petrochemical and oil and gas technological processes," *AIP Conference Proceedings*, vol. 2007, no. 1, p. 050002, 2018. doi: 10.1063/1.5051946.
\bibitem{b4}W. R. Patterson and D. W. Kuthy, "Energy markets," in *Energy Transition and Economics*, 1st ed., Elsevier, 2023, pp. 23-41. doi: 10.1016/b978-0-323-85525-9.00003-9.
\bibitem{b5} G. Faiella, *The Technology of Mesopotamia (The Technology of the Ancient World)*, 1st ed., New York, NY: Rosen Publishing Group, 2005.
\bibitem{b6} P. R. S. Moorey, *Ancient Mesopotamian Materials and Industries: The Archaeological Evidence*, 1st ed., Winona Lake, IN: Eisenbrauns, 1999. 
\bibitem{b7} History of Information, "Oliver Evans Builds the First Automated Flour Mill: Origins of the Integrated and Automated Factory," \url{https://www.historyofinformation.com/detail.php?id=3126} (accedido: 18 de enero de 2025). 
\bibitem{b8} "Evans' Hopper Boy," Wikimedia Commons, \url{https://en.wikipedia.org/wiki/Oliver_Evans#/media/File:Evans'_Hopper}.
\bibitem{b9} R. F. Hirsh, *Technology and Transformation in the American Electric Utility Industry*, 1st ed., Cambridge, UK: Cambridge University Press, 1989.
\bibitem{b10} "The Relay," Technics History, \url{https://technicshistory.com/2017/01/29/the-relay/} (accedido: 18 de enero de 2025).
\bibitem{b11} Encyclopaedia Britannica, "Industrial Revolution," Encyclopaedia Britannica, \url{https://www.britannica.com/event/Industrial-Revolution} (accedido: 18 de enero de 2025).
\bibitem{b12}Petrochemical Complexes in North America," Offshore Technology, \url{https://www.offshore-technology.com} (accedido: 18 de enero de 2025). 
\bibitem{b13} GlobalData, "Petrochemicals Database," GlobalData Marketplace, \url{https://www.globaldata.com/marketplace/oil-and-gas/petrochemicals/} (accedido: 18 de enero de 2025).
\bibitem{b14}Mordor Intelligence, "Global Digital Transformation Market," Mordor Intelligence, \url{https://www.mordorintelligence.com/es/industry-reports/global-digital-transformation-market} (accedido: 18 de enero de 2025).
\bibitem{b15}Encyclopaedia Britannica, "Bhopal disaster," Encyclopaedia Britannica, \url{https://www.britannica.com/event/Bhopal-disaster} (accedido: 18 de enero de 2025).
\bibitem{b16}K. Bloch and B. K. Vaughen, "Looking Back at the Phillips 66 explosion in Pasadena, Texas: 30 years later," Hydrocarbon Processing, Center for Chemical Process Safety, AIChE, Oct. 2018.
\bibitem{b17}"Pemex's Deer Park oil refinery scales back operations after fatal accident," Reuters, \url{https://www.reuters.com/business/energy/pemexs-deer-park-oil-refinery-scales-back-operations-after-fatal-accident-2024-10-12/} (accedido: 18 de enero de 2025).
\bibitem{b18}U.S. Department of Labor, Occupational Safety and Health Administration, "Process Safety Management of Highly Hazardous Chemicals," OSHA Standard 29 CFR 1910.119, 2024. [Online]. Available: \url{https://www.osha.gov/laws-regs/regulations/standardnumber/1910/1910.119}. [Accedido: 18 de enero de 2025].
\bibitem{b19}"Directive 2014/34/EU of the European Parliament and of the Council of 26 February 2014 on the harmonisation of the laws of the Member States relating to equipment and protective systems intended for use in potentially explosive atmospheres," Official Journal of the European Union, L96, pp. 309–356, Mar. 29, 2014.
\bibitem{b20}"IEC 60079-0:2022, Explosive atmospheres – Part 0: Equipment – General requirements," International Electrotechnical Commission, Geneva, Switzerland, 9th ed., 2022.
\bibitem{b21}"UL 1203: Explosion-Proof and Dust-Ignition-Proof Electrical Equipment for Use in Hazardous (Classified) Locations," Underwriters Laboratories, Northbrook, IL, USA, 6th ed., 2013.
\bibitem{b22}"IEC 60529:1989+A1:1999+A2:2013, Degrees of protection provided by enclosures (IP Code)," International Electrotechnical Commission, Geneva, Switzerland, 2nd ed., 2013.
\bibitem{b23}"NEMA 250-2020, Enclosures for Electrical Equipment (1000 Volts Maximum)," National Electrical Manufacturers Association, Rosslyn, VA, USA, 2020.
\bibitem{b24}Wang, Xiaodong., Wang, Yutao., Yang, Xiaoshuai. (2019). National-secret-based offline industrial control software authentication method.  
\bibitem{b25}L. Greenemeier, "Cybersecurity threats to critical infrastructure are growing, and Triton malware is a prime example," *MIT Technology Review*, Mar. 5, 2019. [Online]. Available: \url{https://www.technologyreview.com/2019/03/05/103328/cybersecurity-critical-infrastructure-triton-malware/}. [Accessed: Jan. 21, 2025].
\bibitem{b26}MIT Technology Review en español, "Así se propaga Triton, el 'malware' que amenaza la industria mundial," *MIT Technology Review en español*, Feb. 12, 2019. [Online]. Available: \url{https://www.technologyreview.es/s/11009/asi-se-propaga-triton-el-malware-que-amenaza-la-industria-mundial}. [Accessed: Jan. 21, 2025].
\bibitem{b27}Statista, "La industria del petróleo en el mundo," *Statista*, [Online]. Available: \url{https://es.statista.com/temas/9767/la-industria-del-petroleo-en-el-mundo/}. [Accessed: Jan. 21, 2025
\bibitem{b28}Macrotrends, "Ganancias Netas Anuales," [Online]. Available: \url{https://www.macrotrends.net/}. [Accessed: Jan. 21, 2025].
\bibitem{29}A. C. Arinze, O. A. Ajala, C. C. Okoye, O. C. Ofodile, and A. I. Daraojimba, "Evaluating the integration of advanced IT solutions for emission reduction in the oil and gas sector," Environmental Science and Technology Journal, vol. 5, no. 3, 2024. [Online]. Available: \url{https://doi.org/10.51594/estj.v5i3.862}.
\end{thebibliography}
\end{document}